\documentclass[%
 aip,
 amsmath,amssymb,
 reprint,%
]{revtex4-1}

\usepackage{graphicx}
\usepackage{dcolumn}
\usepackage{bm}
\usepackage{floatrow}
\usepackage{hyperref}
\usepackage{float}
\usepackage{multirow}

\usepackage[utf8]{inputenc}
\usepackage[T1]{fontenc}
\usepackage{mathptmx}
\usepackage{etoolbox}

\makeatletter
\def\@email#1#2{%
 \endgroup
 \patchcmd{\titleblock@produce}
  {\frontmatter@RRAPformat}
  {\frontmatter@RRAPformat{\produce@RRAP{*#1\href{mailto:#2}{#2}}}\frontmatter@RRAPformat}
  {}{}
}%
\makeatother
\begin{document}

\preprint{AIP/123-QED}
\title{Auxiliary Physics-Informed Neural Networks for Forward, Inverse, and Coupled Radiative Transfer Problems}
\author{R. Riganti}
 \affiliation{Department of Physics, Boston University, 590 Commonwealth Avenue, Boston, Massachusetts 02215, USA}
\author{L. Dal Negro}%
 \email{dalnegro@bu.edu}
\affiliation{Department of Physics, Boston University, 590 Commonwealth Avenue, Boston, Massachusetts 02215, USA}
\affiliation{Department of Electrical $\And$ Computer Engineering, and Photonics Center, Boston University, 8 Saint Mary’s Street, Boston, Massachusetts 02215, USA
}%

\affiliation{Division of Materials Science $\And$ Engineering, 
Boston University, 15 St. Mary’s street, Brookline, MA 02446,USA}

\date{\today}

\begin{abstract}
In this paper, we develop and employ auxiliary physics-informed neural networks (APINNs) to solve forward, inverse, and coupled integro-differential problems of radiative transfer theory (RTE). Specifically, by focusing on the relevant slab geometry and scattering media described by different types of phase functions, we show how the proposed APINN framework enables the efficient solution of Boltzmann-type transport equations through multi-output neural networks with multiple auxiliary variables associated to the Legendre expansion terms of the considered phase functions. Furthermore, we demonstrate the successful application of APINN to the coupled radiation-conduction problem of a participating medium and find distinctive temperature profiles beyond the Fourier thermal conduction limit. Finally, we solve the inverse problem for the Schwarzschild-Milne integral equation and retrieve the single scattering albedo based solely on the knowledge of boundary data, similar to what is often available in experimental settings. The present work significantly expands the current capabilities of physics-informed neural networks for radiative transfer problems that are relevant to the design and understanding of complex scattering media and photonic structures with applications to metamaterials, biomedical imaging, thermal transport, and semiconductor device modeling.
\end{abstract}
\maketitle

\section{Introduction}
Over the past few years, there has been a growing interest in developing deep learning (DL) and artificial intelligence (AI) algorithms for electromagnetic wave engineering, metamaterials design, and radiative transport problems\cite{schmidt_recent_2019, wei_deep-learning_2019, jiang_deep_2021}. Rapidly emerging approaches include training artificial neural networks (ANNs) to solve complex inverse problems, parameter estimation in structured photonic environments, and in strongly scattering media\cite{sanghvi_embedding_2020,liu_seagle_2018,kamilov_learning_2015,molesky_inverse_2018, ma_deep_2021}. Although successfully demonstrated with respect to several inverse design problems, traditional methods remain essentially data-driven techniques and require time-consuming training steps and massive datasets\cite{mehta_high-bias_2019, rudy_data-driven_2017}. In order to improve on purely data-driven methods, it is essential to constrain and regularize them by leveraging the underlying physics of the investigated problems, thus relaxing the burden on training and data acquisition. Building on the firm foundation of the universal approximation theorem for multi-layer ANNs\cite{barron_universal_1993, goodfellow_deep_2016}, physics-informed neural networks (PINNs) have recently emerged as a powerful framework for the efficient solution of both forward and inverse problems mathematically described by partial differential equations (PDEs) of integer or fractional orders\cite{pang_fpinns_2019,karniadakis_physics-informed_2021,lu_deepxde_2021, raissi_physics-informed_2019}. The approach of PINNs has been successfully applied to a number of differential problems in engineering ranging from Navier-Stokes
fluid dynamics, solid mechanics, and thermal transport\cite{cai_physics-informed_2021,mao_physics-informed_2020,gigli_predicting_2023}. Moreover, PINNs have shown remarkable results and noise robustness in the solution of electromagnetic inverse problems for metamaterials design, radiative transfer, imaging, and in the parameter retrieval of resonant photonic nanostructures\cite{wang_understanding_2021,chen_physics-informed_2022, chen_physics-informed_2020, lu_physics-informed_2021}. However, the solution of Boltzmann-type, integro-differential transport equations using PINNs still poses significant challenges due to the need to resort to numerical quadrature methods such as Gauss-Legendre or Gauss-Chebyshev for the approximation of the integral terms\cite{lakshmikantham_theory_1995}. Such methods add computational complexity and inevitably introduce quadrature errors in the numerical solutions\cite{li_physics-informed_2021,li_physics-informed_2022,mishra_physics_2021}.

In order to eliminate such problems, a new PINN framework called auxiliary physics-informed neural networks (APINNs) was recently introduced by Yuan et al.\cite{yuan_-pinn_2022}. This approach allows one to recast integro-differential equations into equivalent differential ones through the introduction of a network architecture containing additional auxiliary variables at its output, each corresponding to an integral term in the original, constrained by suitable relations. Therefore, the APINN formulation avoids the numerical approximation of integrals that are instead directly "guessed" by the network at a minimal cost, significantly improving both the numerical accuracy and computational efficiency compared to traditional PINNs.

In this paper, we develop a general APINN framework for solving relevant forward and inverse integro-differential transport equations of radiative transfer theory, which is a domain of vital importance in science and engineering with applications to complex photonic devices, medical imaging, metamaterials,  thermal transport, as well as astrophysics, climate dynamics, and nuclear engineering\cite{howell_thermal_2020, modest_radiative_2021, mishchenko_multiple_2017}.
In particular, we address and demonstrate APINN formulations for the accurate solution of forward, inverse, and coupled radiation-conduction problems of radiative transport in the relevant slab geometry for different choices of scattering phase functions. 

Our paper is organized as follows: in Section~\ref{specs}, we will provide a brief introduction to the radiative transfer equation (RTE), along with a description of the general APINN employed throughout this paper. In Section~\ref{results}, we discuss forward problems for different phase functions governing the scattering processes. Specifically, we present benchmarked solutions for isotropic, Rayleigh, and Henyey-Greenstein scattering phase functions that are often utilized in engineering applications \cite{modest_radiative_2021, howell_thermal_2020, wang_biomedical_2007}. In Section~\ref{coupled}, we discuss the APINN solution of a coupled radiation-conduction problem, enabling the accurate description of radiation transfer in a partecipating medium. Lastly, in Section~\ref{inverse}, we show the solution of a canonical inverse problem described by the Schwarzschild-Milne integral equation, and we show that the radiative intensity solution and the single scattering parameters are accurately retrieved solely based on intensity data at the boundaries of the slab. 

Our work shows that APINNs possess the flexibility, accuracy, and robustness required to become a powerful tool for inverse scattering and thermal transport modeling beyond the limitations of Fourier theory. Therefore, this work expands significantly upon the current capabilities and range of applications of PINNs methods and paves the way to the study of higher-dimensional transport problems in strongly scattering media with applications to nanophotonics, metamaterials, biomedical imaging, and optoelectronic device modeling.

\section{APINNs for radiative transfer problems} \label{specs}
The framework of radiative transfer theory for the study of complex scattering media was originally developed in astrophysics as a way to quantitatively describe the radiative equilibrium in interstellar clouds, planetary and stellar atmospheres\cite{chandrasekhar_radiative_2016}. Radiative transfer theory has found a very wide range of applications beyond astrophysics, including biomedical optics\cite{wang_biomedical_2007}, atmospheric science\cite{mishchenko_multiple_2017}, radiation hydrodynamics\cite{castor_radiation_2004,pomraning_equations_2005} and remote sensing\cite{tsang_scattering_2000,ishimaru_wave_1978}. For example, the propagation of light through fogs and clouds, white paints or paper, milky and turbid liquids, human tissue, and the brain can be adequately described by the classical theory of radiation transfer that we discuss in this paper using APINNs.
The radiation transfer theory is founded upon the RTE, which is a Boltzmann-type integro-differential equation expressing the detailed energy balance for the propagation of directed energy ﬂow, or radiance, through a multiply scattering discrete random medium. For scalar waves in three spatial dimensions the RTE can be written as follows:
\begin{equation}
\begin{split}
    \frac{1}{c}\frac{\partial I(\bm{r},\hat{s},t)}{\partial t}&{}=-\hat{s}\cdot\nabla I(\bm{r},\hat{s},t)-(\kappa+\sigma) I(\bm{r},\hat{s},t)+\\
    &{}\sigma\int_{4\pi}I(\bm{r},\hat{s}',t)p(\hat{s}',\hat{s})d\Omega'+S(\bm{r},\hat{s},t)
\end{split}
\label{eq:rtespatial}
\end{equation}
where $\kappa$ and $\sigma$ are the absorption and scattering coefficients, respectively.
Here $S(\bm{r},\hat{s},t)$ denotes a generic source term and $p(\hat{s}',\hat{s})$ is the phase function describing the angular distribution of the scattering process.
Alternatively, after introducing the optical thickness $\tau$ and the single scattering albedo $\omega$ as: 
\begin{align}
    \tau(S) = \int_{S'=0}^{S}\beta(S')dS'&=\int_{S'=0}^{S}[\kappa(S')+\sigma(S')]dS'\\
    \omega = \frac{\sigma}{\beta}&=\frac{\sigma}{\kappa+\sigma}
\end{align}
one can rewrite Eq.~\ref{eq:rtespatial} in the alternative form:
\begin{equation}
\begin{split}
    \frac{1}{\beta c}\frac{\partial I(\bm{\tau},\hat{s},t)}{\partial t}&{}=-\hat{s}\cdot\nabla_\tau\, I(\bm{\tau},\hat{s},t)-I(\bm{\tau},\hat{s},t)+\\
    &{}\omega\int_{4\pi}I(\bm{\tau},\hat{s}',t)p(\hat{s}',\hat{s})d\Omega'+S(\bm{\tau},\hat{s},t)
\end{split}
\label{eq:rteopt}
\end{equation}
which is the RTE in its standard form. For a detailed discussion and derivation of the RTE, we refer the reader to references~\onlinecite{chandrasekhar_radiative_2016,howell_thermal_2020,modest_radiative_2021}. In essence, the RTE states that a directed beam of light in a uniform random medium loses energy through divergence and extinction, including both absorption and scattering away from the beam (i.e., out-scattering contributions), and it gains energy from radiation sources, ﬂuorescence or scattering events that redirect it towards the beam (i.e., in-scattering contributions). In the standard formulation, wave interference effects, polarization and non-linearity in the medium are neglected. Radiative transfer theories for vector waves have also been developed but are outside the scope of this work and more details on these subjects can be found in references \onlinecite{mishchenko_multiple_2017, ishimaru_wave_1978}. Even for the relevant slab geometry, the RTE introduced above is generally very difficult to solve\cite{frisch_radiative_2022}. Analytic solutions only exist for very simple cases while in many realistic situations, numerical methods such as Monte Carlo transport simulations are usually employed\cite{graham_stochastic_2013}. For this reason, the RTE is often approximated, under suitable conditions, by the simpler but less accurate diffusion equation\cite{wang_biomedical_2007}.

\begin{figure}
\includegraphics[scale=0.525]{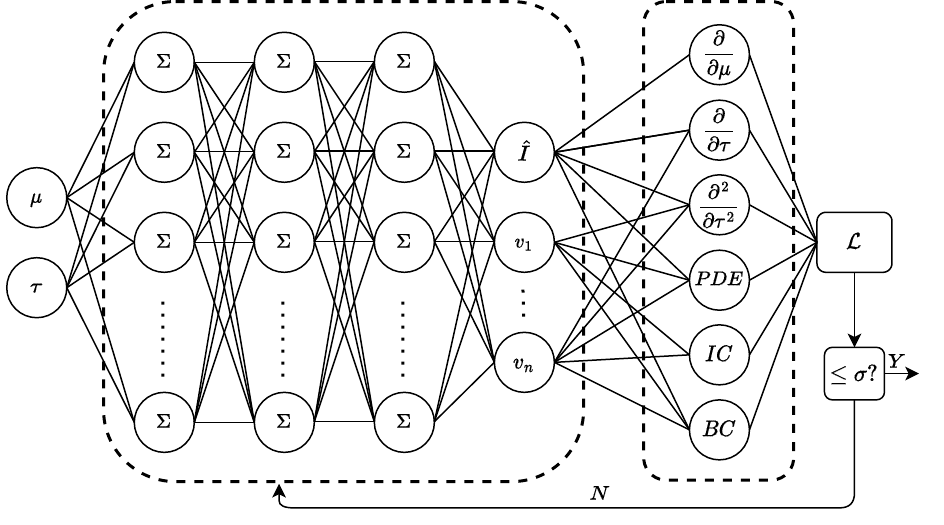}
\caption{\label{fig:apinn} Schematics of the APINN solving the RTE problem in a slab. The FCNN has $N+1$ outputs where $N$ is the number of auxiliary variables $v_i$ required in the Legendre polynomial expansion of the scattering phase function. The outputs of the network are then used to satisfy the PDE, initial conditions (ICs), and boundary conditions (BCs) of the differential equation, which are then combined into the loss function $\mathcal{L}$. During the training process, the loss function is minimized until it passes a threshold $\sigma$.}
\end{figure}

\begin{table*}
\caption{\label{tab:net} Summary of the general information of the APINN framework for the numerical experiments that will follow. The number of neurons listed is for each layer of the APINN, and we used 80 neurons per layer only for the Henyey-Greenstein phase function and coupled problem due to the greater number of auxiliary variables in the network architecture and complexity of the problem, respectively.}
\begin{ruledtabular}
\renewcommand{\arraystretch}{1.25}
\begin{tabular}{cccccccddc}
Solver& Learning rate & Regularization & Epochs & Activation Function&Layers&Neurons& $$\mathcal{N}_{int}$$ & $$\mathcal{N}_{b}$$ & GPU\\
\hline
L-BFGS-B & 1e-5 & L2, $\lambda=0.001$ & 7500 & tanh & 10 & 40, 80 & 8192 & 2048 & V100 32GB\\
\end{tabular}
\end{ruledtabular}
\end{table*}

In our paper, we developed APINNs to obtain the forward and inverse solution of the scalar RTE in the steady-state and for different choices of phase functions. However, the developed framework can be naturally extended to time-dependent and vector RTE problems, anisotropic phase functions, and arbitrary nonlinear responses. All the implementations of the APINN algorithms developed in this paper are obtained in the powerful TensorFlow environment\cite{tensorflow2015-whitepaper}.

The general APINN network utilized to solve forward and inverse RTE problems in the slab geometry is illustrated in Fig.~\ref{fig:apinn}. We considered a fully connected neural network (FCNN) with input vector $\bm{x}=(\bm{\tau}, \bm{\mu})$ with randomly distributed values of the optical thickness $\tau$ and $\mu=\cos\theta$ over a two-dimensional spatial-angular domain $\Omega$ and output that is the predicted surrogate  $\hat{I}(\mu,\tau;\bm{\tilde{\theta}})$ of the RTE solution $I(\mu,\tau;\bm{\theta})$. Here, $\theta$ denotes the angle of the directed energy flow with the axis $z$ perpendicular slab's surface and $\bm{\tilde{\theta}}$ is the vector of weights and biases of our FCNN. In addition, the FCNN outputs $n$ auxiliary variables $v_{i}(\mu,\tau;\bm{\tilde{\theta}})$, each corresponding to an integral expansion term in the RTE.
The outputs of the APINN are then used to compute, by means of automatic differentiation (AD), the derivatives of $\hat{I}(\mu,\tau;\bm{\tilde{\theta}})$ and $v_{i}(\mu,\tau;\bm{\tilde{\theta}})$, along with the PDE, initial conditions, and boundary conditions, depending on the nature of the problem. Each calculated value is then combined into a term of the loss function $\mathcal{L}(\bm{\tilde{\theta}})$ defined as:
\begin{equation}
    \begin{split}
        \mathcal{L}(\bm{\tilde{\theta}}) = &{} \mathcal{L}_{int}(\bm{\tilde{\theta}};\mathcal{N}_{int})+\mathcal{L}_{b}(\bm{\tilde{\theta}};\mathcal{N}_{b})\\
        &{}+ \mathcal{L}_{aux}(\bm{\tilde{\theta}};\mathcal{N}_{aux})+\lambda\sum_i\bm{\tilde{\theta}}_{i}^{2}
    \end{split}
    \label{eq:loss}
\end{equation}
In the expression above,  
\begin{align}
    \mathcal{L}_{int}(\bm{\tilde{\theta}};\mathcal{N}_{int}) =\frac{1}{|\mathcal{N}_{int}|}\sum\nolimits_{\bm{x}\in\mathcal{N}_{int}} \left|\left| f\left( \bm{x};\hat{I},\frac{\partial\hat{I}}{\partial \tau},v_0,\dots, v_n \right) \right|\right|^2
\end{align}
denotes the loss term calculated in the interior domain $\Omega$ and
\begin{align}
    \mathcal{L}_{b}(\bm{\tilde{\theta}};\mathcal{N}_{b}) = \frac{1}{|\mathcal{N}_{b}|}\sum\nolimits_{\bm{x}\in\mathcal{N}_{b}} \left|\left| \mathcal{B}(\hat{I},\bm{x}) \right|\right|^2
\end{align}
is the loss term for the boundary conditions of the RTE where $\bm{x}\in \partial\Omega$. Moreover,
\begin{align}
    \mathcal{L}_{aux}(\bm{\tilde{\theta}};\mathcal{N}_{aux}) = \frac{1}{|\mathcal{N}_{aux}|}\sum\nolimits_{\bm{x}\in\mathcal{N}_{aux}} \left|\left| f\left( \bm{x};\frac{\partial v_0}{\partial \mu},\dots, \frac{\partial v_n}{\partial \mu} \right) \right|\right|^2
\end{align}
denotes the loss term associated to the auxiliary conditions that define the APINN model. $\mathcal{N}_{int},\,\mathcal{N}_{b},\,\mathcal{N}_{aux}$ denote the number of residual points for each loss term, and the last term in Eq.~\ref{eq:loss} is an L2 regularization included in our simulations to avoid overfitting during training\cite{goodfellow_deep_2016}.

Table~\ref{tab:net} summarizes the training and APINN network parameters for the simulations studied throughout this paper. In the forward simulations of Section~\ref{results}, we decided to analyze RTE problems in the slab geometry with different scattering phase functions of ever-increasing terms in the Legendre series expansion, resulting in an increasing number of integrals in the RTE, while keeping the general network and training parameters the same. The Legendre series expansion of the RTE phase function will be discussed in detail in Section~\ref{rayleigh}. We thus start from the Schwarzschild-Milne equation, whose RTE has only one integral, and its corresponding APINN requires only one auxiliary variable. Then, we study the RTE with the Rayleigh phase function, whose Legendre expansion has two non-zero terms, resulting in two auxiliary outputs in the network. Finally, we study the Henyey-Greenstein (HG) phase function, whose series expansion was truncated at the tenth term, introducing ten auxiliary variables in the APINN. This approach allowed us to present a reliable scaling analysis when APINN is employed to solve integro-differential problems with kernels whose series expansions converge at different speeds. In the next section, we start presenting our APINN results, and we begin by addressing the Schwarzschild-Milne equation in a slab.

\section{Results and Discussion} 
\subsection{Solutions of forward problems in a slab}\label{results}

\begin{figure*}
\includegraphics[scale=0.3]{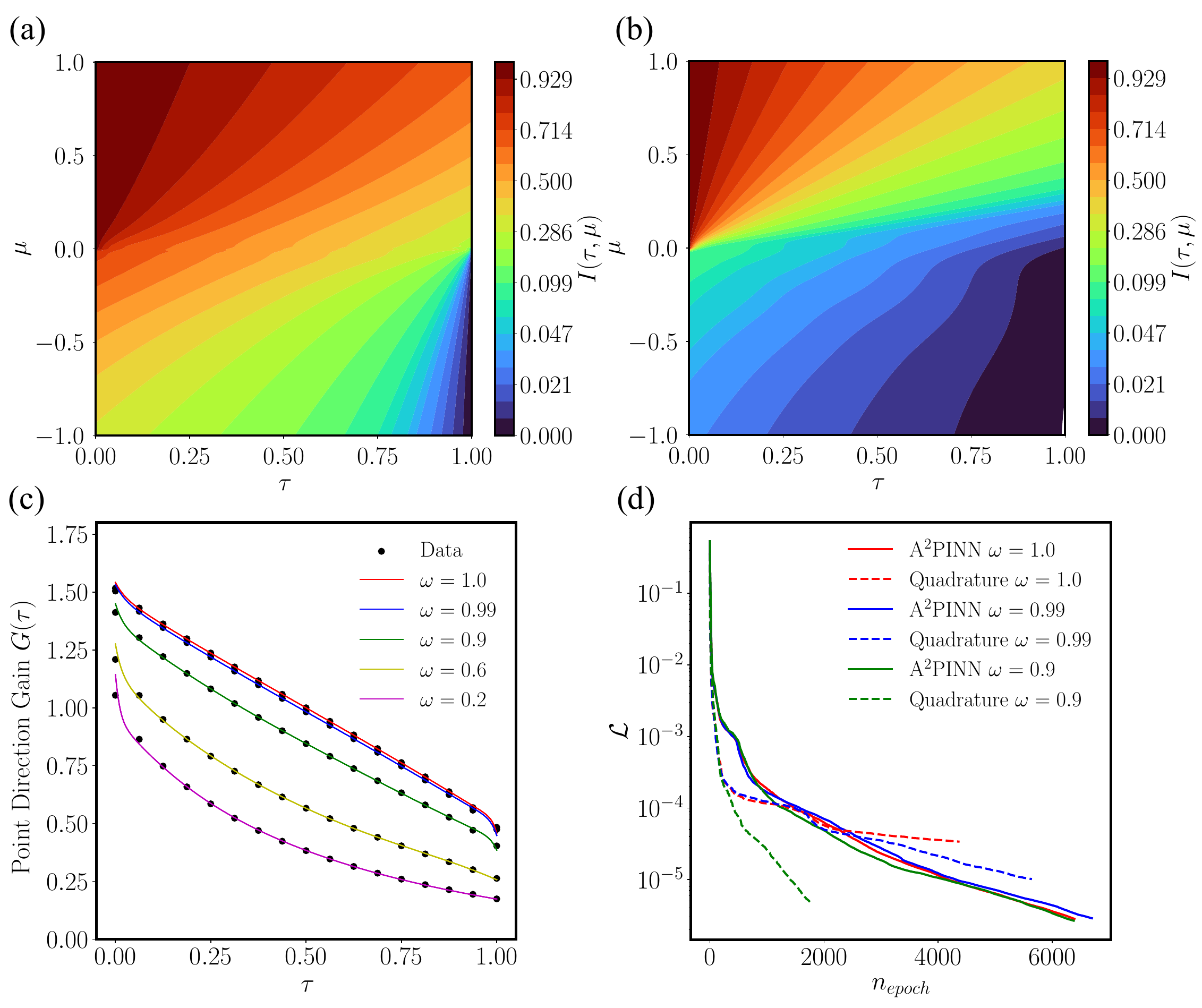}
\caption{\label{fig:milne} (a) and (b) APINN solutions for the Milne equation in a slab, with the single scattering albedo set to $\omega=1$ and $\omega=0.2$, respectively. $\omega$ sets the strength of the scattering term in the PDE. (c) Validation of the APINN solutions using Van de Hulst's data results for the point direction gain $G(\tau)$. (d) Comparison of the APINN training performance in minimizing the objective loss function with the quadrature solution: the APINN method consistently trains independently of scattering strength, whereas the quadrature method's performance worsens as the scattering strength, represented by $\omega$, increases.}
\end{figure*}

\subsubsection{The Schwarzschild-Milne equation}

We first consider the time-independent radiative transfer problem in a slab governed by the RTE. As discussed by Howell\cite{howell_thermal_2020}, this steady-state condition of the RTE is valid under the assumption that the radiation intensity is unaffected by photon time-of-flight effects, reducing Eq.~\ref{eq:rteopt} to the form investigated here:
\begin{equation}
    \mu \frac{dI(\tau,\mu)}{d\tau} + I(\tau,\mu)=\frac{\omega}{2}\int_{-1}^{1} I(\tau,\mu')\Phi(\mu,\mu')d\mu' 
\label{eq:rte}
\end{equation}
When $\Phi(\mu,\mu')=1$, the equation above becomes the well-known Schwarzschild-Milne integral equation describing isotropic scattering processes. 
The corresponding boundary conditions are\cite{van_rossum_multiple_1999}:
\begin{subequations}
    \begin{gather}
        I(0, \mu) = I_{0},\;0<\mu<1\\
        I(\tau_{0},\mu) = 0,\; -1<\mu<0
    \end{gather}
\label{eq:rtebc}
\end{subequations}
In order to solve the Schwarzschild-Milne integral equation using the APINN framework, we recast it into an equivalent differential problem introducing the auxiliary variable $v(\mu;\tau)$, which is constrained by the following system:
\begin{subequations}
    \begin{gather}
        \mu\frac{dI}{d\tau}+I-\frac{\omega}{2}v(1)=0\\
        \begin{split}
            &{}v(\mu;\tau)=\int_{-1}^{\mu}I(\mu';\tau)d\mu'\\
            &{}v(-1;\tau)=0, \quad \frac{dv}{d\mu}(\mu;\tau)=I(\tau,\mu)
        \end{split}
    \end{gather}
    \label{eq:milne}
\end{subequations}
We then train the APINN to solve the problem for different values of the albedo $\omega$ varying from $0.2$ to $1.0$. Table~\ref{tab:milne} shows the speed and accuracy of our APINN implementation in solving the Milne problem. In the large scattering limit of $\omega \geq 0.9$, APINN minimized the loss function with values that are two orders of magnitude lower and for a fraction of the time than for the equivalent geometry displayed in Ref.~\onlinecite{mishra_physics_2021}, where a quadrature method was employed. Two representative APINN solutions for the spatial-angular distributions of the radiation intensity for $\tau_{\text{max}}=1.0$ are displayed in Fig.~\ref{fig:milne} (a) and (b). To benchmark our solutions using the tables calculated by Van de Hulst's in Ref.~\onlinecite{hulst_multiple_1656}, we computed the zeroth moment or point-direction gain $G(\tau)$ of the radiative intensity, which is defined as\cite{modest_radiative_2021,howell_thermal_2020,chandrasekhar_radiative_2016, hulst_multiple_1656}:
\begin{equation}
    G(\tau) = \int_{-1}^1 I(\tau,\mu) d\mu 
\end{equation}
Fig.~\ref{fig:milne} (c) displays the validation data of $G(\tau)$ calculated by Van de Hulst and the solution from our network, showing an excellent agreement achieved by the APINN framework. This is further confirmed by the average relative error between the two solutions displayed in the last column of Table~\ref{tab:milne}. Fig.~\ref{fig:milne} (d) shows a comparison between the APINN and the standard PINN quadrature loss function to solve the same problem, as implemented in  Ref.~\onlinecite{mishra_physics_2021}. In this figure, we display the loss function versus the number of epochs for the three largest scattering values of $\omega$. We can immediately notice that the quadrature solution is heavily affected in its performance by the scattering strength, and the L-BFGS-B solver terminates the training early because the loss function has already saturated to its minimum value and is not decreasing further. In contrast, the APINN's loss function monotonically decreases independently of $\omega$. This result confirms the robustness, flexibility, and accuracy of the APINN framework in solving transport problems for strongly scattering media. In a variety of engineering applications, however, the material's response is not isotropic. Therefore, in Section~\ref{rayleigh}, we employ the APINN framework to solve the RTE in a slab with an anisotropic Rayleigh scattering phase function.  

\begin{table}
\caption{\label{tab:milne} APINN training information for the Milne equation in a slab. The last column shows the average relative error between the network solution of $G(\tau)$ and Van de Hulst's results.}
\begin{ruledtabular}
\renewcommand{\arraystretch}{1.25}
\begin{tabular}{cccc}
$\omega$ & Training time & Loss & Avg. Rel. Error \\
\hline
0.2 & 3 min & $7\times10^{-6}$ & $7\times10^{-2}$\\
0.6 & 10 min & $5\times10^{-6}$ & $8\times10^{-2}$\\
0.9 & 10 min & $3\times10^{-6}$ & $7\times10^{-2}$\\
0.99 & 10 min & $3\times10^{-6}$ & $7\times10^{-2}$\\
1.0 & 10 min & $3\times10^{-6}$ & $7\times10^{-2}$\\
\end{tabular}
\end{ruledtabular}
\end{table}

\subsubsection{The Rayleigh scattering phase function}\label{rayleigh}
The Rayleigh phase function is employed to study anisotropic light scattering processes in various fields, from optics to astronomy\cite{chandrasekhar_radiative_2016}. The phase function reads:
\begin{equation}
    p(\text{cos}\theta) = \frac{3}{4}(1+\text{cos}\theta^2)
\end{equation}
and because the scattering from spherically symmetric particles is cylindrically symmetric with respect to the incoming direction, this symmetry holds after averaging over all possible orientations. Therefore, in these situations, the phase function depends on $\phi-\phi'$ and one can compute this average resulting in the projected phase function\cite{van_rossum_multiple_1999}:
\begin{equation}
    p_0(\mu,\mu')=\int\frac{d\phi}{2\pi}\frac{d\phi'}{2\pi}p(\mu,\phi;\mu',\phi')
\end{equation}
Using the equality $\mu=\cos\Theta=\bm{n}\cdot\bm{n}'=\text{sin}\theta\text{sin}\theta'\text{cos}(\phi-\phi')+\text{cos}\theta\text{cos}\theta'$ one obtains:
\begin{equation}
    p_0(\mu,\mu')=\frac{3}{8}(3-\mu^2-\mu'^2+3\mu^2\mu'^2)
\end{equation}
To facilitate the calculations and the auxiliary variable formulation of the APINN framework, one typically considers the expansion of the scattering phase function in Legendre polynomials:
\begin{equation}
    \Phi(\mu,\mu') = \sum_{\ell=0}^{\infty}w_{\ell}P_{\ell}(\mu)P_{\ell}(\mu')
\label{eq:legendre}
\end{equation}
Note that, for the Rayleigh phase function, the only nonzero $w_\ell$ terms are $w_0=1.0$ and $w_2=0.1$. Therefore, Eq.~\ref{eq:rte} in a slab with Rayleigh scattering becomes
\begin{equation}
    \mu \frac{dI(\tau,\mu)}{d\tau} + I(\tau,\mu)=\frac{\omega}{2}\int_{-1}^{1}I(\tau,\mu')\sum_{\ell=0}^{\infty}w_{\ell}P_{\ell}(\mu)P_{\ell}(\mu')d\mu' 
\end{equation}
and after rearranging terms and truncating the series expansion at $\ell=2$ we get:
\begin{equation}
\begin{split}
    \mu \frac{dI(\tau,\mu)}{d\tau} + I(\tau,\mu)=\frac{\omega}{2}&{}\Big[w_{0}P_{0}(\mu)\int_{-1}^{1}I(\tau,\mu')P_{0}(\mu')d\mu'\\
    &{}+w_{2}P_{2}(\mu)\int_{-1}^{1}I(\tau,\mu')P_{2}(\mu')d\mu'\Big]
\end{split}
\end{equation}
Finally, we recast the problem by adding two auxiliary variables to the network with their respective constraints as follows:
\begin{subequations}
    \begin{gather}
        \mu \frac{dI(\tau,\mu)}{d\tau} + I(\tau,\mu)=\frac{\omega}{2}\left[w_{0}P_{0}(\mu)v_{0}(1)+w_{2}P_{2}(\mu)v_{2}(1)\right] \label{eq:rayrte} \\
        \begin{split}
            &{}v_{0}(\mu;\tau)=\int_{-1}^{\mu}I(\tau,\mu')P_{0}(\mu')d\mu'\\
            &{}v_{0}(-1;\tau)=0, \quad \frac{dv_{0}}{d\mu}(\mu;\tau)=I(\tau,\mu)P_{0}(\mu)\\
        \end{split}\\
        \begin{split}
            &{}v_{2}(\mu;\tau)=\int_{-1}^{\mu}I(\tau,\mu')P_{2}(\mu')d\mu'\\
            &{}v_{2}(-1;\tau)=0, \quad \frac{dv_{2}}{d\mu}(\mu;\tau)=I(\tau,\mu)P_{2}(\mu)
        \end{split}
    \end{gather}
\end{subequations}

\begin{figure*}
\includegraphics[scale=0.35]{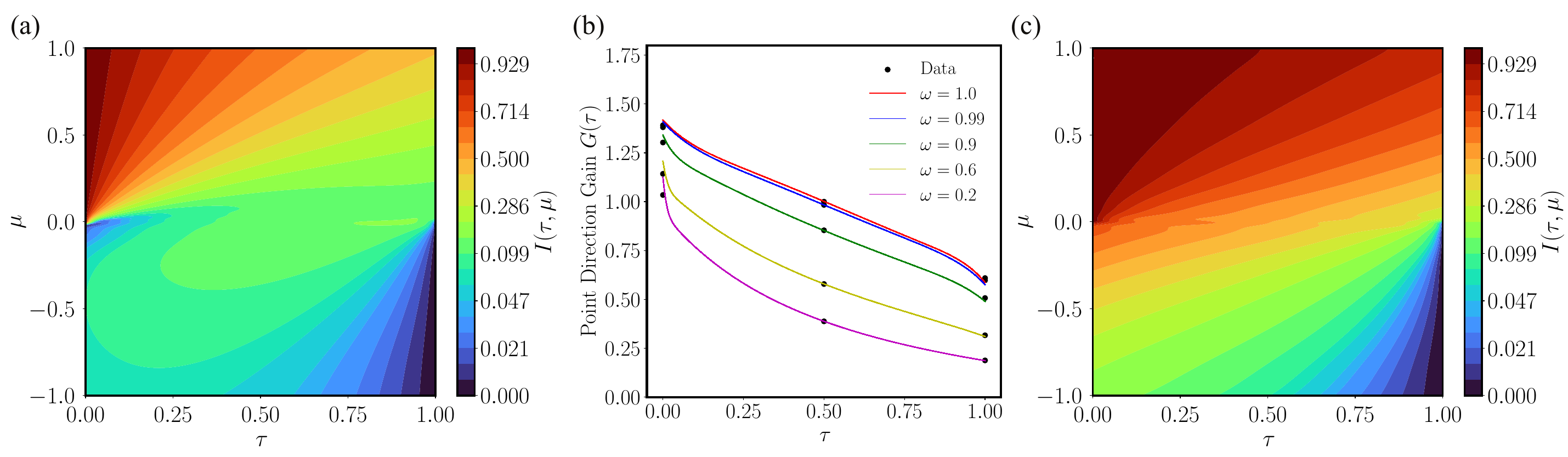}
\caption{\label{fig:hgray} (a) APINN solution for the RTE with Rayleigh scattering phase function, where the single scattering albedo $\omega(\tau)$ depends on the optical thickness. (b) Validation of the APINN solutions for the HG phase function with $g=0.5$ using Van de Hulst's data results for the point direction gain $G(\tau)$. (c) Representative APINN solution for the RTE in a slab with HG scattering phase function, $g=0.5$, $\omega=1.0$.}
\end{figure*}

Due to the lack of benchmark solutions for Rayleigh scattering in a slab, we decided to consider a physical system similar to the one studied by Mishra and Molinaro in Ref.~\onlinecite{mishra_physics_2021}, namely the case where the single scattering albedo depends on the optical thickness $\tau$ of the material. In this case, Eq.~\ref{eq:rayrte} becomes
\begin{equation}
    \begin{split}
            \frac{dI(\tau,\mu)}{d\tau} + I(\tau,\mu)=\frac{\omega(\tau)}{2}\Big[&{}w_{0}P_{0}(\mu)v_{0}(1;\tau)\\
            &{}+w_{2}P_{2}(\mu)v_{2}(1;\tau)\Big]
    \end{split}
\end{equation}
To solve this problem, we train APINN with the parameters specified in Table~\ref{tab:net}, using 40 neurons per layer. The training for this solution took 12 minutes, and the final value of the loss function $\mathcal{L}$ was $10^{-6}$, demonstrating the adaptivity and flexibility of APINN in solving anisotropic scattering problems. Fig.~\ref{fig:hgray}(a) displays the APINN radiative intensity solution as a function of $\mu$ and the optical thickness. This result highlights the flexibility of APINN in finding the solution to an analytically intractable problem\cite{mishra_physics_2021}. In turn, this motivates us to study the RTE with strongly anisotropic scattering properties modeled by the Henyey-Greenstein (HG) phase function. 

\subsubsection{The Henyey-Greenstein phase function}
Here we consider the forward RTE problem in the slab with the Henyey-Greenstein (HG) phase function governing the scattering processes. The HG phase function finds applications in astrophysics, atmospheric optics, and biomedical imaging, and it depends on both the cosine of the incident angle and the asymmetry factor $g\in[0,1]$ that appears in the equation below\cite{modest_radiative_2021, howell_thermal_2020, abdoulaev_optical_2005, arridge_optical_2009,bal_inverse_2009}:
\begin{equation}
    p(\mu,g) = \frac{1-g^2}{(1-2g\mu+g^2)^{3/2}}
\end{equation}
where $\mu = \text{cos}\theta$. In the limit of $g\rightarrow0$, the HG phase function reduces to isotropic scattering, while in the limit of $g\rightarrow1$, HG describes strongly anisotropic scattering events.\\ 

\begin{table}
\caption{\label{tab:hg} APINN training information for the HG phase function in a slab, $g=0.5$. The last column shows the average relative error between the network solution of $G(\tau)$ and Van de Hulst's table.}
\begin{ruledtabular}
\renewcommand{\arraystretch}{1.25}
\begin{tabular}{cccc}
$\omega$ & Training time & Loss & Avg. Rel. Error \\
\hline
0.2 & 49 min & $8\times10^{-6}$ & $3\times10^{-2}$\\
0.6 & 50 min & $2\times10^{-5}$ & $2\times10^{-2}$\\
0.9 & 65 min & $3\times10^{-5}$ & $6\times10^{-3}$\\
0.99 & 65 min & $6\times10^{-5}$ & $2\times10^{-3}$\\
1.0 & 65 min & $5\times10^{-5}$ & $1\times10^{-3}$
\end{tabular}
\end{ruledtabular}
\end{table}

As for the Rayleigh phase function, the HG phase function can be rewritten using the Legendre polynomials expansion in Eq.~(\ref{eq:legendre}). However, unlike the Rayleigh case, the Legendre expansion converges more slowly, and additional terms need to be included to achieve accurate numerical results:
\begin{equation}
    \begin{split}
        \mu \frac{dI(\tau,\mu)}{d\tau} + I(\tau,\mu)=\frac{\omega}{2}\Big[&{}w_{0}P_{0}(\mu)\int_{-1}^{1}I(\tau,\mu')P_{0}(\mu')d\mu'\\
        &{}+w_{1}P_{1}(\mu)\int_{-1}^{1}I(\tau,\mu')P_{1}(\mu')d\mu'\\
        &{}+\dots\\
        &{}+w_{n}P_{n}(\mu)\int_{-1}^{1}I(\tau,\mu')P_{n}(\mu')d\mu'\Big]
    \end{split}
\end{equation}
where:
\begin{equation}
    w_\ell = (2n+1)g^n
\end{equation}

In our numerical studies, we chose to benchmark the RTE with HG phase function, $g=0.5$, which allowed us to utilize  Van de Hulst's tables as validation data\cite{hulst_multiple_1980}. The polynomial expansion of the phase function was truncated after ten terms, introducing ten auxiliary variables and their corresponding constraint conditions in the simulation:
\begin{subequations}
\begin{gather}
\begin{split}
    &{}\mu \frac{dI(\tau,\mu)}{d\tau} + I(\tau,\mu)=\frac{\omega}{2}\Big[w_{0}P_{0}(\mu)v_{0}(1;\tau)\\
    &{}+\dots+w_{10}P_{10}(\mu)v_{10}(1;\tau)\Big]
\end{split}\\
\begin{split}
    &{}v_{0}(\mu;\tau)=\int_{-1}^{\mu}I(\tau,\mu')P_{0}(\mu')d\mu'\\
    &{}v_{0}(-1;\tau)=0, \quad \frac{dv_{0}}{d\mu}(\mu;\tau)=I(\tau,\mu)P_{0}(\mu)
\end{split}\\
    \dots \nonumber \\
\begin{split}
    &{}v_{10}(\mu;\tau)=\int_{-1}^{\mu}I(\tau,\mu')P_{10}(\mu')d\mu'\\
    &{}v_{10}(-1;\tau)=0, \quad \frac{dv_{10}}{d\mu}(\mu;\tau)=I(\tau,\mu)P_{10}(\mu)
\end{split}
\end{gather}
\label{eq:hgrte}
\end{subequations}
Table~\ref{tab:hg} provides a summary of the APINN training for this problem. Considering the larger number of auxiliary variables, we trained using 80 neurons per layer instead of 40. Similarly to the isotropic and Rayleigh cases, the loss function is minimized to extremely low values with a minor trade-off in speed due to the larger number of auxiliary variables in the system, as the second and third columns of  Table~\ref{tab:hg} demonstrate. The accuracy of these results, displayed in the last column of  Table~\ref{tab:hg}, confirms the versatility of the APINN framework, which excels in solving even strong anisotropic scattering problems. Fig.~\ref{fig:hgray} (c) shows a representative solution of the radiation intensity when $\omega=1.0$, and Fig.~\ref{fig:hgray} (b) displays the benchmarked solutions for this problem by comparing the integrated radiative intensity $G(\tau)$ calculated from the APINN network with the Van de Hulst's data. These results open the doors for multiple biomedical, metamaterials, and nano-optics applications where the HG phase function is often utilized to model realistic scattering processes\cite{abdoulaev_optical_2005, arridge_optical_2009, bal_inverse_2009}.

\subsection{The coupled radiation-conduction problem of a participating medium}\label{coupled}
\begin{figure}
\includegraphics[scale=0.525]{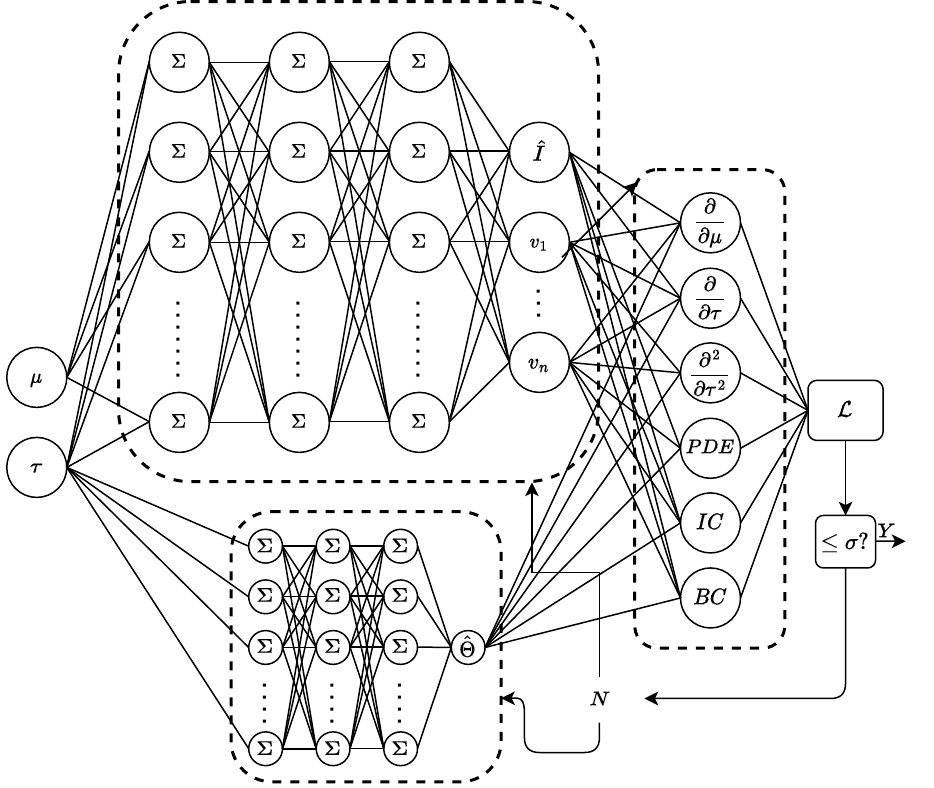}
\caption{\label{fig:capinn} Schematics of the APINN solving the coupled radiation-conduction problem in a slab. Here, two FCNNs representing the intensity of radiation $\hat{I}(\mu,\tau)$ and temperature $\hat{\Theta}(\tau)$ are combined in a loss function that minimizes the PDE, initial conditions (ICs), and boundary conditions (BCs) of two coupled partial differential equations. The training process minimizes the loss function $\mathcal{L}$ until it passes a threshold $\sigma$.}
\end{figure}
We now apply our APINN method to the solution of a coupled problem in radiative transfer theory. Specifically, we consider a conducting and participating slab that couples to the radiation hitting the boundary in the steady-state. Such problems have been analyzed extensively in the literature\cite{molesky_inverse_2018, howell_thermal_2020, slimi_transient_2004, tong_multiscale_2021, liu_non-fourier_2001,gotz_coupling_2002,moura_neto_introduction_2013,larsen_simplified_2002,klar_boundary_1998,ghattassi_galerkin_2016,ismail_gray_2006,modest_elliptic_2008}, but, to our knowledge, have never been solved using physics-informed neural networks. Here, we use the APINN framework to analyze this problem, where the slab's temperature profile is governed by a Poisson-like equation with a coupling term to the RTE\cite{moura_neto_introduction_2013}. We will further analyze how the conduction-radiation parameter $N_{CR}$ affects the traditional Fourier temperature solution in the steady-state when significant temperature differences are imposed at the boundaries of the slab. The conduction-radiation parameter $N_{CR}$ measures the ratio of conductive to radiative heat contributions in a given medium, and it is defined as\cite{howell_thermal_2020}:
\begin{equation}
    N_{CR}=\frac{k \beta}{4 k_B T^3} = \frac{k (\kappa + \sigma)}{4 k_B T^3}
\label{eq:N}
\end{equation}
For the simulations that follow, we chose to study coupled systems where $N_{CR}$ varies from 10 (for $N \rightarrow \infty$, we get the Fourier limit) to 0.001 (for $N \rightarrow 0$, radiative processes dominate).

We consider the heat transfer problem due to conduction and radiation in a participating medium presented by Ref.~\onlinecite{moura_neto_introduction_2013} governed by the two following coupled integro-differential equations:
\begin{align}
    &{}\frac{d^2\Theta}{d\tau^2}-\frac{(1-\omega)}{N_{CR}}\left[ 
    \Theta^4(\tau)-\frac{1}{2}G(\tau)\right]=0\\
    &{}\mu \frac{dI(\tau,\mu)}{d\tau} + I(\tau,\mu)=H[\Theta(\tau)]+\frac{\omega}{2}\int_{-1}^{1} I(\tau,\mu')\Phi(\mu,\mu')d\mu'\\
    &{}\text{for}\quad0<\tau<1,\quad -1\leq\mu\leq1,\quad \Phi(\mu,\mu')=1, \quad \omega=0.9 \nonumber
\end{align}
where the temperature is being modeled by the normalized adimensional quantity $\Theta=T/T_1$. The coupling terms are:
\begin{equation}
    G(\tau)=\int^{1}_{-1}I(\tau,\mu)d\mu,\quad H[\Theta(\tau)]=(1-\omega)\Theta^4
\end{equation}
and $G(\tau)$ is the zeroth moment of the intensity 
$I(\tau,\mu)$. The boundary conditions are:
\begin{align}
    &{} I(0,\mu)=1, \mu\in(0,1], \quad\text{and}\quad I(1,\mu)=0, \mu\in[-1,0)\\
    &{} \Theta(0)=1 \quad\text{and}\quad \Theta(1)=T_2/T_1
\end{align}

\begin{figure}
\includegraphics[scale=0.31]{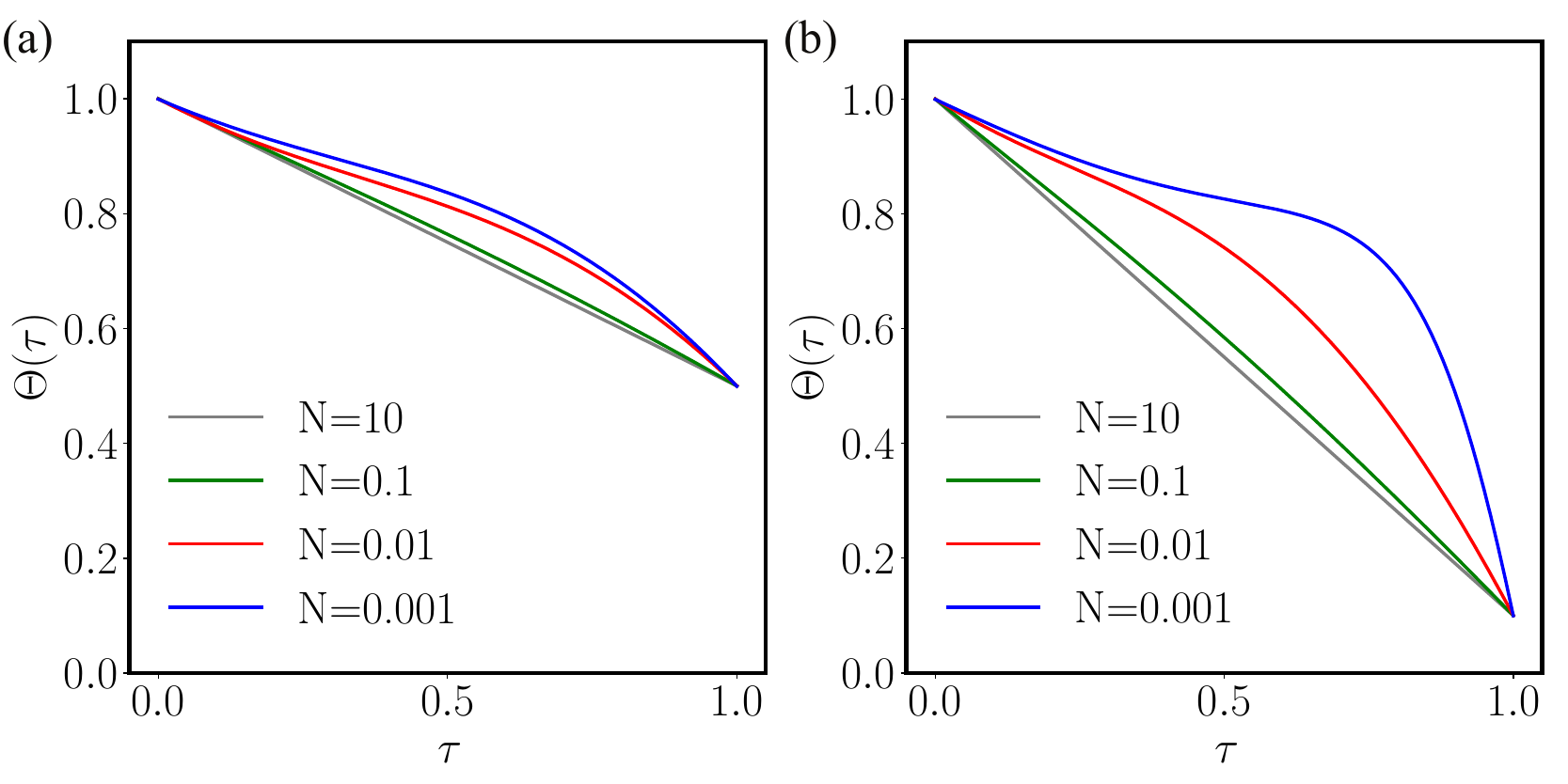}
\caption{\label{fig:coupled} (a) and (b) APINN results displaying the temperature profile in the conductive participating medium of Eq.~\ref{eq:coupled} for a temperature jump of 150K and 270K, respectively. The simulations have been conducted for four sets of values of the conduction-radiation parameter $N_{CR}$: as $N_{CR}$ decreases, the solution changes from the well-known Fourier temperature profile to a beyond-Fourier solution characterized by a dominant scattering contribution.}
\end{figure}
\begin{table}
\caption{\label{tab:coupled} APINN Training information for radiative-conductive coupled problem in a slab.}
\begin{ruledtabular}
\renewcommand{\arraystretch}{1.25}
\begin{tabular}{ccccc}
$\Delta\Theta$ & $N_{CR}$ & Training time & $\mathcal{L}_{\hat{I}(\tau,\mu)}$ & $\mathcal{L}_{\hat{\Theta}(\tau)}$ \\
\hline
\multirow{4}{*}{150K} &   10 & 18 min & $6\times10^{-4}$ & $6\times10^{-6}$\\
 &   0.1 & 18 min & $5\times10^{-4}$ & $5\times10^{-7}$\\
 &   0.01 & 18 min & $5\times10^{-4}$ & $7\times10^{-6}$\\
 &   0.001 & 18 min & $5\times10^{-4}$ & $8\times10^{-5}$\\
 \hline 
\multirow{4}{*}{270K} &   10 & 33 min & $5\times10^{-4}$ & $2\times10^{-5}$\\
 &   0.1 & 30 min & $5\times10^{-4}$ & $2\times10^{-6}$\\
 &   0.01 & 31 min & $5\times10^{-4}$ & $6\times10^{-6}$\\
 &   0.001 & 30 min & $5\times10^{-4}$ & $4\times10^{-5}$
\end{tabular}
\end{ruledtabular}
\end{table}
Since the problem involves two undetermined coupled functions, we modified the architecture of the APINN framework. The changes are illustrated in Fig.~\ref{fig:capinn}: the input parameters are passed to the radiative intensity network $\hat{I}(\tau,\mu)$ with auxiliary variables as for the uncoupled cases discussed so far, but the spatial variable $\tau$ is also used to train simultaneously the adimensional temperature network $\hat{\Theta}(\tau)$. The coupled problem recasted in the APINN formalism reads: 
\begin{align}
     \frac{d^2\Theta}{d\tau^2}-\frac{(1-\omega)}{N_{CR}}\left[ 
    \Theta^4(\tau)-\frac{1}{2}v(1;\tau)\right]=0\\
    \mu \frac{dI(\tau,\mu)}{d\tau} + I(\tau,\mu)=H[\Theta(\tau)]+\frac{\omega}{2}v(1;\tau)
\end{align}
where we introduced the auxiliary variable $v(\mu;\tau)$ and its corresponding conditions like in Eq.~(\ref{eq:milne}):
\begin{subequations}
\begin{gather}
    v(\mu;\tau)=\int_{-1}^{\mu}I(\mu';\tau)d\mu',\\
    v(-1;\tau)=0, \quad \frac{dv}{d\mu}(\mu;\tau)=I(\tau,\mu)
\end{gather}   
\label{eq:coupled} 
\end{subequations}
By means of automatic differentiation, the outputs of the two networks are then used to compute the required PDE conditions, initial conditions, and boundary conditions, which are then incorporated into the coupled loss function:
\begin{align}
    \mathcal{L}= \mathcal{L}_{\hat{I}(\tau,\mu)} + \mathcal{L}_{\hat{\Theta}(\tau)}
\end{align}
To solve this problem, we coupled the APINN network for the radiative intensity with a PINN estimating the dimensionless temperature $\Theta(\tau)$, with parameters according to Table~\ref{tab:net}. Fig.~\ref{fig:coupled} shows the solutions for the coupled problem when two different temperature jumps are imposed at the rightmost boundary. Fig.~\ref{fig:coupled}(a) displays a  $\Delta\Theta=150K$, whereas Fig.~\ref{fig:coupled}(b) a $\Delta\Theta=270K$. Moreover, we analyze the dimensionless temperature behavior when the conduction-radiation parameter $N_{CR}$ decreases, as previously investigated in Refs. \onlinecite{modest_radiative_2021, howell_thermal_2020}. It is important to realize that both panels in Fig.~\ref{fig:coupled} display a beyond-Fourier behavior as $N_{CR}$ decreases, demonstrating that the temperature profile is significantly affected by radiative scattering phenomena. Lastly, Table~\ref{tab:coupled} presents some relevant information regarding the APINN training. We note that, even for the coupled case, the APINN successfully minimizes both the temperature loss function $\mathcal{L}_{\hat{\Theta}(\tau)}$ and the radiative intensity loss function $\mathcal{L}_{\hat{\Theta}(\tau)}$ independently of the parameter $N_{CR}$.

\subsection{Inverse problem: retrieval of the albedo from the boundary data}\label{inverse}
Finally, we present here the solution of an inverse problem of radiative transfer theory where we employ APINN to retrieve simultaneously the forward solution of the intensity $I(\tau,\mu)$ and the single scattering albedo $\omega$. We do not, however, introduce synthetic data everywhere in the domain, as it has been done previously in the literature\cite{mishra_physics_2021}, but we limit ourselves to introducing two data points representing the integrated intensity $G(\tau)$ at the edges of the slab, simulating a lab environment with two detectors capturing integrated radiation entering and exiting the slab, respectively. The reason to present an inverse problem in such a fashion is to demonstrate the full potential and capabilities of physics-informed neural networks that, with no additional overhead and computing power, can solve a forward and parameter retrieval problem simultaneously. We thus modify the Schwarzschild-Milne equation for a slab discussed in an earlier section. In particular, Eq.~(\ref{eq:milne}) is changed to include the unknown albedo parameter $\omega_\theta$:
\begin{equation}
    \mu\frac{dI}{d\tau}+I-\frac{\omega_\theta}{2}v(1)=0
\end{equation}

\begin{figure}
\includegraphics[scale=0.4]{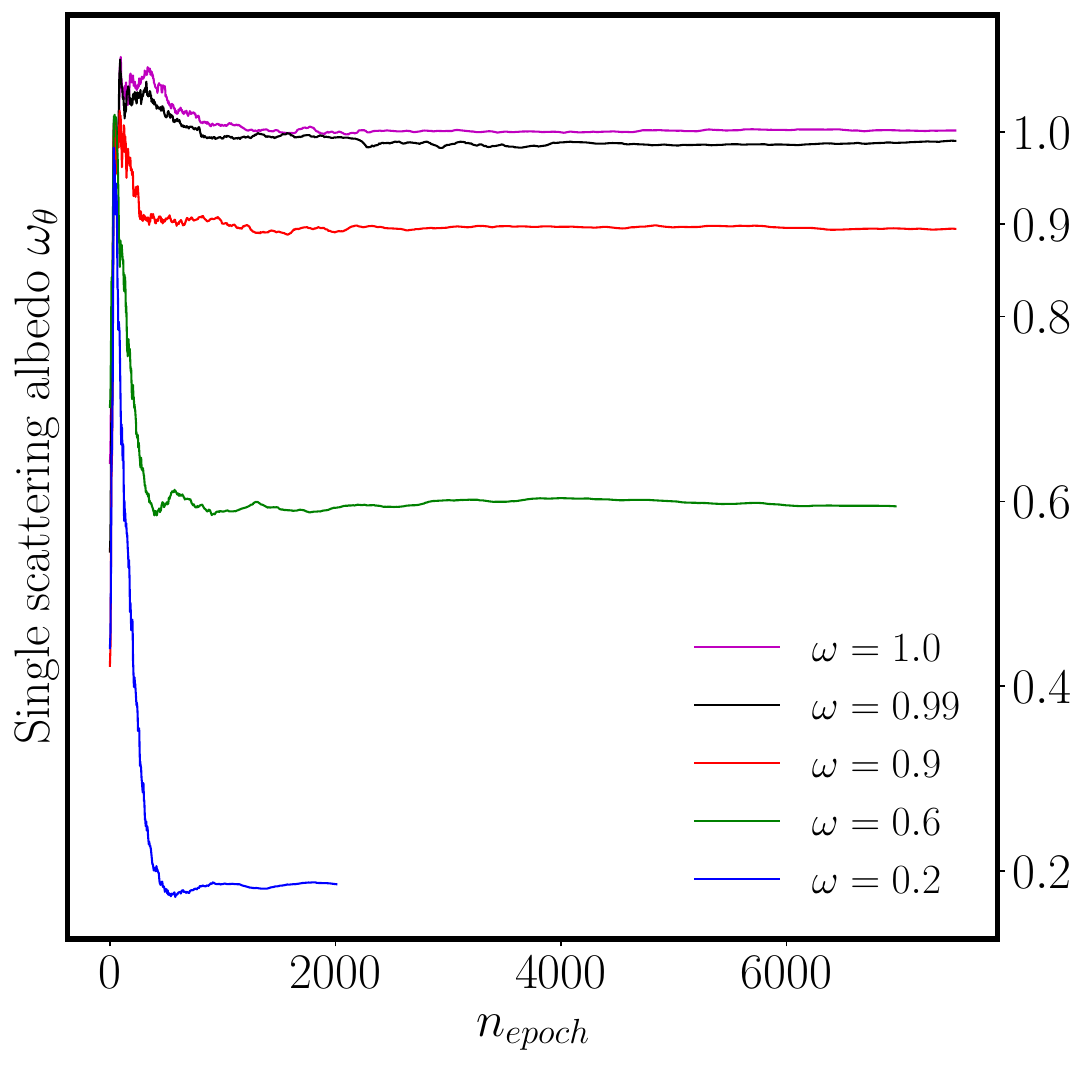}
\caption{\label{fig:inverse} Results showing the convergence of each value of different trainable variables $\omega_\theta$ for separate APINN simulations, each of which where fed different boundary values $G(\tau)$ for the inverse problem, while all the other parameters and PDE conditions stayed the same. As the picture shows, the neural network provided very accurate results across all simulations, training $\omega_\theta$ to the true albedo value $\omega$, especially for high scattering values.}
\end{figure}

and the loss function in Eq.~(\ref{eq:loss}) is modified to include the two synthetic detector data points at the boundaries of the slab:
\begin{equation}
\begin{split}
    \mathcal{L}(\bm{\theta},\bm{\omega}_\theta) &{}= \mathcal{L}_{int}(\bm{\theta},\bm{\omega}_\theta;\mathcal{N}_{int})+\mathcal{L}_{b}(\bm{\theta},\bm{\omega}_\theta;\mathcal{N}_{b})\\
    &{}+\mathcal{L}_{aux}(\bm{\theta},\bm{\omega}_\theta;\mathcal{N}_{aux}) +\mathcal{L}_{inv}(\bm{\theta},\bm{\omega}_\theta;\mathcal{N}_{inv})
\end{split}
\end{equation}
where
\begin{equation}
\begin{split}
    &{}\mathcal{L}_{inv}(\bm{\theta},\bm{\omega}_\theta;\mathcal{N}_{inv})= \\
    &{} \frac{1}{|\mathcal{N}_{inv}|}\sum_{(\tau,\mu)\in\mathcal{N}_{inv}} \left|\left| \int^{1}_{-1}\hat{I}(\tau,\mu)d\mu - G(\tau)\right|\right|^2=\\
    &{}\frac{1}{2}\left(\left|\left| \int^{1}_{-1}\hat{I}(0,\mu)d\mu - G(0)\right|\right|^2 + \left|\left| \int^{1}_{-1}\hat{I}(1,\mu)d\mu - G(1)\right|\right|^2\right)
\end{split}
\end{equation}

\begin{table}
\caption{\label{tab:inv} APINN training information for the inverse problem. For high scattering values $\omega$, the APINN loss function quickly converges to values below $10^{-5}$. The last column displays the average relative error between the true value of $\omega$ and APINN's $\omega_\theta$.}
\begin{ruledtabular}
\renewcommand{\arraystretch}{1.25}
\begin{tabular}{cccc}
$\omega$ & Training time & Loss & Rel. Error \\
\hline
0.2 & 4 min & $3\times10^{-6}$ & $10^{-2}$\\
0.6 & 16 min & $5\times10^{-6}$ & $5\times10^{-3}$\\
0.9 & 17 min & $5\times10^{-6}$ & $5\times10^{-3}$\\
0.99 & 17 min & $10^{-5}$ & $2\times10^{-4}$\\
1.0 & 17 min & $6\times10^{-6}$ & $10^{-3}$
\end{tabular}
\end{ruledtabular}
\end{table}

Fig.~\ref{fig:inverse} displays the fast convergence of the retrieved APINN parameter $\omega_\theta$ to the actual value $\omega$. Each line corresponds to a different APINN training procedure during which the only data points added were $G(0)$ and $G(1)$ obtained from the Van de Hulst's tables\cite{hulst_multiple_1656}, and used to minimize $\mathcal{L}_{inv}$ during the training process. Despite the loss term with two data points was not weighted differently from the interior or boundary ones, APINN achieved a precise inversion of the parameter of interest. In fact, as displayed in Table~\ref{tab:inv}, the loss function converges independently of the albedo $\omega$ and with great precision, as displayed by the relative error between the known albedo and the predicted APINN albedo $\omega_\theta$ shown in the last column of Table~\ref{tab:inv}. Therefore, APINN retrieved the correct parameter of interest $\omega_\theta$ when only two points were added during the training process.

\section{Conclusions} \label{conclusions}
Throughout this paper, we have described different applications of APINN for solving the radiative transfer equation, which is a Boltzmann-type transport equation. We successfully solved forward problems in a slab with both isotropic and anisotropic scattering phase functions and irrespective of the albedo. The results presented improved upon previous attempts to use physics-informed neural networks for solving the RTE in both accuracy and speed\cite{mishra_physics_2021}. Furthermore, we presented the solution of the first coupled radiation-conduction problem in a participating medium using the APINN framework and we showed that the loss functions of coupled neural networks quickly converged to a low minimum value below $10^{-5}$. Our findings open the possibility to utilize APINN to analyze higher dimensional systems and discover more interesting physics with applications to metamaterials and semiconductor device modeling. Finally, we solved an inverse problem following a setup that replicates an experimental setting with data points at the boundary of the system. It will be interesting in future studies to build on the APINN platform to address higher dimensional coupled, inverse coupled, and strongly scattering forward systems with applications to biomedical imaging, nanophotonics, metamaterials, and thermal modeling of semiconductor devices.

\begin{acknowledgments}
We acknowledge the support from the U.S. Army Research Office, RF-Center managed by Dr. J. Qiu (Grant \#W911NF-22-2-0158). We thank professors Mike Kirby, Akil Narayan, and Shandian Zhe for useful discussions on this topic.
\end{acknowledgments}

\nocite{*}
\bibliography{apinnrte_v7}

\end{document}